\documentclass[sort&compress,preprint,12pt]{elsarticle}
\journal{ }
\usepackage{amsmath,amssymb,graphicx,color,lipsum,mathrsfs}

\begin{document}

\begin{frontmatter}
\title{Instability in Stable Marriage Problem: Matching Unequally Numbered Men and Women}
\author[hyit,unifr]{Gui-Yuan Shi}
\author[hyit,unifr]{Yi-Xiu Kong}
\author[hyit]{Bo-Lun Chen}
\author[shufe]{Guang-Hui Yuan}
\author[unifr]{Rui-Jie Wu\corref{cor1}}\ead{ruijie.wu@unifr.ch}
\cortext[cor1]{Corresponding author}

\address[hyit]{College of Computer Engineering, Huaiyin Institute of Technology, Huaian 233003, PR China
}
\address[unifr]{Department of Physics, University of Fribourg, Fribourg 1700, Switzerland}
\address[shufe]{Fintech Research Institute, Shanghai University of Finance and Economics, Shanghai 200433,  PR China}

\begin{abstract}
The Stable Marriage Problem is to find a one-to-one matching for two equally sized sets of agents. Due to its widespread applications in the real world, especially the unique importance to the centralized match maker, a very large number of questions have been extensively studied in this field. This article considers a generalized form of stable marriage problem, where different numbers of men and women need to be matched pairwise and the emergence of single is inevitable. Theoretical analysis and numerical simulations confirm that even small deviations from equal number of two sides can have a large impact on matching solution of Gale-Shapley Algorithm. These results provide insights to many of the real-world applications when matching two sides with unequal number. 
\end{abstract}

\end{frontmatter}

\section{Introduction}
The Stable Marriage Problem(SMP), as the name suggests, is a study of two groups of men and women, who need to be matched pairwise. In addition, there are also extensive applications in college admissions\cite{Gale1962}, labor markets\cite{Roth1984} and many other social systems\cite{Roth1992}. The recent advances in Internet technology have introduced the stable marriage problem into new application scenarios, for example, assigning a large number of users to internet server\cite{Maggs2015}, matching peers in a P2P network according to preference\cite{Lebedev2007}, the resource allocation in 5G networks\cite{Hasan2015}, and the matching mechanism of real-world dating website\cite{Hitsch2010}, etc. 

Many researchers studied the algorithms and properties on this fascinating and practical problem. The study of the Stable marriage problem originated from 1962, the Gale-Shapley algorithm was proposed\cite{Gale1962}, and the algorithm guarantees that there is always a stable match for an equal number of men and women\cite{Roth1982}. In 2012, the Nobel Prize in Economics was awarded to Lloyd S. Shapley and Alvin E. Roth "for the theory of stable allocations and the practice of market design." In this algorithm, each unengaged man issues proposal to his favorite woman who has not yet rejected him, and the woman comes to decide which one to choose, compares this suitor with the current provisional partner, and retains the one that is more desirable to her. This solution is proved to be one of the stable men-optimal solutions.

Besides mathematicians and computer scientists, statistical physicists are also interested in this issue\cite{Chakraborti2015, Zhang2017}. They have introduced many theoretical methods which is used to deal with the statistical properties of a large number of interacting particles to the Stable Marriage Problem and have made many contributions. The mean field theory was applied to calculate satisfaction of both men and women in the stable marriage problem, and it was found that men, who are the active side, were far happier than women\cite{Omero1997}. This method can also be used to estimate the number of stable solutions\cite{Pittel1989,Dzierzawa2000}. The replica method used in Spin glass is applied to study the global optimal solution in bipartite matching, although this is not a stable solution\cite{Mezard1985,Mezard1987,Dotsenko2000,Shi2016}. In addition, some interesting varied issues have been studied, such as the impact of matching on the matching of partial information\cite{Zhang2001,Laureti2003}, spatial distribution, intrinsic fitness\cite{Caldarelli2001}, and acceptance threshold\cite{Dzierzawa2000}.

So far, main stream research has been focusing on the matching of equal number of the two sides. In fact, however, real-world matching problems are seldom a match of precisely equal size of two sides. We have seen many examples in daily life, the student adminssion, job market, the recent 'One to one Poverty Alleviation Program' policy promoted by the Chinese government, or even the real application of SMP- the marriage matching on dating website\cite{Hitsch2010} etc. These problems are mostly not a matching between equal number of both sides. We are curious to know, whether the original stable solution of Gale-Shapley Algorithm can be applied in these unequal size matching problems, and if not, how would the stable solution of matching be changed? What is the overall happiness of all agents in the stable solution? Many more interesting question may be asked. Here in this paper, we extend the Stable Marriage Problem to a Generalized Stable Marriage Problem(GSMP), which represents a matching between any given size of the two sides. Dzierzawa and Om\'{e}ro\cite{Dzierzawa2000} implemented the numerical simulation to test the matching result of $N+1$ men and $N$ women, and their simulation shows women in this case are far happier than men according to Gale-Shapley Algorithm, which is hugely different from what the original SMP stable solution suggests. This is crucial because it reveals that the Gale-Shapley solution is very sensitive to even the smallest variation in the number of people and actually cannot be directly used to analyze many unequal sized matchign problems. We further thoroughly study the stable solution in GSMP, a matching between any number of men and women. We carry out a theoretical analysis of the stable solution fo GSMP and obtain the average happiness for men and women for any given population, and the result is in perfect match with the numerical simulations.

\section{Method}
The model starts with the scenario that $N$ male and $M$ females need to be matched pairwise. Here we assume that everyone knows all people from the opposite sex, and there is a wish list for each person which represents the ranking of all agents from the other side according to her/his preference. Following many previous research models, a reasonable and simplest assumption is that all wish lists are randomly established and irrelevant. We define an energy function for each agent, which is equal to the ranking of the final partner in their wish list. The lower energy one has, the happier the person is. When $N=M$, it is the conventional bipartite matching problem. Here we extend to the bipartite matching of any size of two sides. When $N \ne M$, obviously there will be some people who will remain single. For these agents, their energy is defined as one worse than the bottom of the wish list, that is to say, the energy is $M+1$ for single men, and $N+1$ for women. Since the number of single persons is obvious, the result of the calculation can be simply converted Defined for other. Here, we use the Greek letter to represent men and English letters to represent women. Their energy is denoted as $e_m$, $e_w$ respectively.

The G-S algorithm runs as follows: unengaged men will continue to send proposals to women, and women keep the one she prefers between suitor and her provisional partner. The process terminates until no man issues proposal again, either all men are engaged, or the unengaged men are rejected by everyone. For $N<=M$, this means that all men are engaged. For the case of $N>M$ , $N-M$ men are rejected by all women, and the rest M men are engaged.

\section{Result and Discussion}
The Stable Marriage Problem of equal size of the two sides has been thoroughly studied by many previous researches. For $M=N$, several studies have proved that in the stable solution of Gale-Sharpley Algorithm, the average energy of women is $\overline{\epsilon_m}=log(N)$, and the average energy of men is $ \overline{\epsilon_w}=N/log(N)$.

\subsection{Matching for $M > N$}

\subsubsection*{(a) The average energy of men}

First, let us consider the situation that number of women $M$ is larger than the number of men $N$. During the process of men proposing to women, we notice that:
1). The total energy of men equals to the number of proposals men have already sent.  
2). Once a woman is engaged, she will remain engaged (perhaps with different men) forever, that is to say the number of partners will never decrease.

Now We focus on the number of matched pairs in the proposing process, if it increases to N, every man has already been assigned with one partner and the proposal process will terminate. When the number of matched pairs is K, there will be M-K women who remain unengaged. If one proposal is sent to an engaged woman, no matter the suitor or the woman’s current provisional partner wins, the number of matched pairs will not change. If a proposal is sent to an unengaged woman, which has a probability of (M-K)/M, the number of matched pairs will increase to K+1. It is easy to know that on average M/(M-K) proposals have to be sent to match one more pair.

Thus, in order to increase the number of matched pairs from 0 to N, the expected number of proposals all men have to send is

\begin{equation}
L_{N,M} =  \sum_{K = 0}^ {N-1} \frac{M}{M-K} =M(\sum_{i = 1}^ {M}\frac{1}{i} - \sum_{j = 1}^ {M-N}\frac{1}{j})
\end{equation}

Hence, the average energy of men when $M>N$ is 

\begin{equation}
\overline{\epsilon_m} = \frac{L_{N,M}}{N} = \frac{M}{N}\mathrm{ln}\frac{M}{M-N}
\end{equation}

\subsubsection*{(b) The average energy of women}

To obtain the average energy of women $\epsilon_m$, let us consider the final stable solution, in which everyone's partner has been determined.
let us assume a woman who was finally paired with the man ranked $\beta$ in her list, so men who ranked higher than $\beta$ in her list did not issue proposal to her. According to the ranking of the men in the women's list, let us denote the men's energy as follows: $\epsilon_{m}(\alpha)$, $\alpha\in\{1,2, \dots,N\}$.
The men who ranked in the top $\beta-1$ in the woman's list must have a better partner than the woman, which means the energy of the woman is higher than that of the assigned partners of these men. This is because if the rank of the lady in a certain men's list is less than the energy value of the him, then he must have already issued a proposal to the woman according to the Gale-Shapley Algorithm, and thus causing a conflict. So we can calculate the probability of a woman matched with her ranking $\beta$ man is:

\begin{equation}
P_\beta=\prod_{\alpha=1}^{\beta-1}(1-\frac{\epsilon_{m}(\alpha)}{M})*\frac{\epsilon_{m}(\beta)}{M}
\end{equation}

For the unengaged women, we denote their energy as $N+1$. These agents have not received any proposal, the probability is:

\begin{equation}
P_{N+1}=\prod_{\alpha=1}^{N}(1-\frac{\epsilon_{m}(\alpha)}{M})
\end{equation}

Similarly, we use $\overline{\epsilon_m}$ to replace $\epsilon_{m}(\alpha)$, then we have: 

\begin{equation}
P_\beta=(1-\frac{\overline{\epsilon_m}}{M})^{\beta-1}*\frac{\overline{\epsilon_m}}{M}
\end{equation}

\begin{equation}
P_{N+1}=(1-\frac{\overline{\epsilon_m}}{M})^{N}
\end{equation}

Consider $\overline{\epsilon_w}=\sum \beta*p_{\beta}$, we can estimate the average energy of women: 

\begin{equation}
\epsilon_w=\frac{\overline{\epsilon_m}}{M}\sum_{i=1}^{N}i*(1-\frac{\overline{\epsilon_m}}{M})^{i-1}+(N+1)*(1-\frac{\overline{\epsilon_m}}{M})^{N}
\end{equation}

The series summation gives: 
\begin{equation}
\epsilon_w=\frac{\overline{\epsilon_m}}{M}\frac{1-(\frac{1-\overline{\epsilon_m}}{M})^N}{(\frac{\overline{\epsilon_m}}{M})^2}+(1-\frac{\overline{\epsilon_m}}{M})^N
\end{equation}

Since we have $\overline{\epsilon_m}=\frac{M}{N}\mathrm{ln}\frac{M}{M-N}$, $(1-\frac{\overline{\epsilon_m}}{M})^N$ approxiemately equals to $\frac{M-N}{M}$

\begin{equation}
\overline{\epsilon_w}=\frac{N}{\overline{\epsilon_m}}+\frac{M-N}{M},
\end{equation}

and also we can see: 
\begin{equation}
\overline{\epsilon_w}*\overline{\epsilon_m}\simeq N
\end{equation}

\subsection{Matching for $M < N$}

\subsubsection*{(a) The average energy of women}

While $M< N$, let us denote the energy of each women as $\epsilon_{i}$, and the average of $\epsilon_{i}$ is denoted by $\epsilon_w$. 

In the final matching state, all the women are matched but $N-M$ men are left single. The probability of a man being single is $\prod_{i=1}^{M}(1-\epsilon_{i}/{N})$, i.e., he ranks worse than any of the current partner in women's lists. In total we have $N-M$ men who are single, which means the probability of being single is $(N-M)/N$, so we have: 

\begin{equation}
\prod_{i=1}^{M}(1-\frac{\epsilon_{i}}{N})=\frac{N-M}{N}
\end{equation}

Approxiemately, we use $\overline{\epsilon_w}$ to replace $\epsilon_{i}$,
\begin{equation}
(1-\frac{\overline{\epsilon_w}}{N})^M=\frac{N-M}{N}
\end{equation}
take the logarithm of both sides we have:

\begin{equation}
\overline{\epsilon_w}=\frac{N}{M}\mathrm{ln}\frac{N}{N-M}
\end{equation}

\subsubsection*{(b) The average energy of men}

Let us consider the final stable state of matching which we know the exact matching result of everyone. For a man $j$, we denote the energy of women in his list as $\epsilon_{w}(j),j\in\{1,2,\dots,M\}$. Then we have the probability that he was accepted by the woman ranked $i^{th}$ in his list: 
\begin{equation}
Q_i=\prod_{j=1}^{i-1}(1-\frac{\epsilon_{w}(j)}{N})\frac{\epsilon_{w}(i)}{N}
\end{equation}

And the probability that he was rejected by all woman is (we assume the energy of finally single man is $M+1$):
\begin{equation}
Q_{M+1}=\prod_{j=1}^{M}(1-\frac{\epsilon_{w}(j)}{N})
\end{equation}

We can approxiemate $\epsilon_{w}(j)$ with the mean value $\overline{\epsilon_w}$, for N is large. By averaging over $i$ we obtain: 
\begin{equation}
\overline{\epsilon_m}=\sum_{i=1}^{M}i(1-\frac{\overline{\epsilon_w}}{N})^{i-1}\frac{\overline{\epsilon_w}}{N}+(M+1)(1-\frac{\overline{\epsilon_w}}{N})^M
\end{equation}

\begin{equation}
\overline{\epsilon_m}=\frac{\overline{\epsilon_w}}{N}\frac{1-(1-\frac{\overline{\epsilon_w}}{N})^M}{(\frac{\overline{\epsilon_w}}{N})^2}+(1-\frac{\overline{\epsilon_w}}{N})^M
\end{equation}
since $(1-{\overline{\epsilon_w}}/{N})^M={(N-M)}/{N}$, we have: 

\begin{equation}
\overline{\epsilon_w}(\overline{\epsilon_m}-\frac{N-M}{M})=M
\end{equation}
\subsection{Numerical Simulations}
\begin{figure}[htb]
\center\scalebox{0.6}[0.6]{\rotatebox{0}{\includegraphics{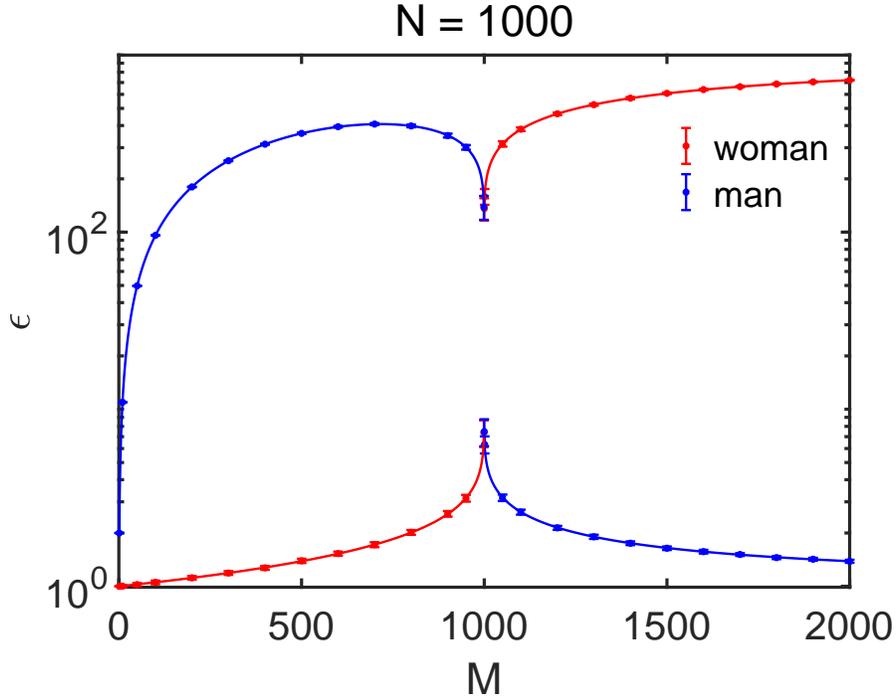}}}
\caption{
The average energy of $N$ men and $M$ women versus the number of women $M$, $N$ is set to 1000. The result is obtained by averaging over 100 realizations. The dots and their error bars represent the simulations result and the standard deviation, the solid lines indicate our theoretical predictions.  
}
\label{fig1}
\end{figure}

The simualtion result is shown in Fig.1. Without loss of generality, we fix the number of men equal to 1000 and vary the number of women $M$ from 1 to 2000. Under our definition of single's energy, as our theoretical result predicted,  the average energy of men increases in the beninning and almost saturate when $M$ reaches around 700. The energy plateau is mainly caused by our definition that single men's energy is $M+1$. Although other definition may apply, we choose this definition out of many reasonable ones because it facilitates our theoretical analysis. It is notable that the average energy is low when $M$ is close to 1. It is insturctive to many social phenomena. When woman are too few($M$ close to 1 ), most of men end up single in the matching. But one's feeling of happiness, as written by the sociologist Ruut Veenhoven\cite{Veenhoven1991},"...happiness results from comparison...", may come from the relative comparison among surroudings. 

The energy of both men and women have dramatic changes when $M$ approaches to 1000. The energy of men has a sharp drop in this region while the energy of women experiences a big rise. The energy of active side and passive side exchange their positions. As $M$ increses, the energy of men continues to decrease, at the meantime, the energy of women keeps growing which is natural for the situation that more women competes for a certain number of men. In total, our simulation result verified our previous theoretical analysis.

For many of the actual matching problems, the change in the number of people's happiness is not as sensitive to the theoretical predictions. This may be due to some intrinsic factors, such as appearance, test scores, university rankings, work ability, salary level, etc., which will cause the wish list to become relevant. One of our studies recently reveals that the correlation of the people’s wish lists can significantly reduce the inherent instability of Gale-Shapley Algorithm on unequal numbers of people.

\section{Conclusion}
In summary, we extend the study of conventional bipartite matching problem to unequal numbers of two sides. This advancement has a realistic impact because the numbers of matching parties are often different in many scenarios in the real world, and the losers of the competition are widespread. For the traditional N male N female matching problem, it is widely accepted that the Gale-Shapley algorithm leads to a matching result in which the active side occupies a huge advantage. However, we find that even by reducing only one woman, the men of the active side become obviously disadvantageous, which means that original stable matching solution is super sensitive for changes in the number of matching members. 

In this paper we thoroughly study the matching solution of unequally siezed Stable Mariiage Problem, and provide both theoretical solution and numerical simulations. These discoveries help us to further understand the structure and properties of the SMP solution, it also sheds a light on the matching when match makers or resource allocators who need to deal with many of the real bipartite matching problems with scarsity of one side.

\section*{Conflict of Interest}

The authors declare no conflict of interest.

\section*{Acknowledgements}

We would like to thank Prof. Yi-Cheng Zhang and Wen-Yao Zhang for helpful discussions. This research is supported in part by the Chinese National Natural Science Foundation under grant No. 61602202 and Natural Science Foundation of Jiangsu Province under grants No. BK20160428, BK20161302. GYS and YXK acknowleges the support from China Scholarship Council (CSC).

\end{document}